\documentclass[aps,prl,groupaddress,superscriptaddress,reprint,amsmath,amssymb,amsfonts,graphicx]{revtex4-1}
\usepackage{graphicx}
\usepackage{physics}
\usepackage{bm}
\usepackage{bbold}
\usepackage{color}
\usepackage{xcolor}

\begin{document}
\preprint{APS/123-QED}

\newcommand{\state}[1]{p_{#1,\boldsymbol{x}}(t)}
\newcommand{\avg}[1]{\langle #1 \rangle}

\newcommand{\add}[1]{\textcolor{black}{#1}}

\title{\color{black}Excitation-inhibition balance controls information encoding in neural populations}

\author{Giacomo Barzon}
\affiliation{Padova Neuroscience Center, University of Padova, Padova, Italy}

\author{Daniel Maria Busiello}
\thanks{D.M.B. and G.N. contributed equally to this work, and are listed alphabetically.}
\affiliation{Max Planck Institute for the Physics of Complex Systems, Dresden, Germany}

\author{Giorgio Nicoletti}
\thanks{D.M.B. and G.N. contributed equally to this work, and are listed alphabetically.}
\affiliation{ECHO Laboratory, École Polytechnique Fédérale de Lausanne, Lausanne, Switzerland}

\begin{abstract}
\noindent Understanding how the complex connectivity structure of the brain shapes its information-processing capabilities is a long-standing question. By focusing on a paradigmatic architecture, we study how the neural activity of excitatory and inhibitory populations encodes information on external signals. We show that \add{at long times} information is maximized at the edge of stability, where \add{inhibition balances excitation, both in linear and nonlinear regimes. In the presence of multiple external signals,} this maximum corresponds to the entropy of the \add{input} dynamics. By analyzing the case of a prolonged stimulus, we find that stronger inhibition is \add{instead} needed to maximize the instantaneous sensitivity, revealing an intrinsic trade-off between short-time responses and long-time accuracy. In agreement with recent experimental findings, our results \add{pave the way} for a \add{deeper} information-theoretic understanding of how \add{the balance between excitation and inhibitions controls optimal} information-processing \add{in neural populations}.
\end{abstract}

\maketitle

\noindent From sensory perception to task-driven behaviors and decision-making processes, the brain constantly receives and integrates large amounts of environmental information. Both the encoding and processing of information in the cortex involve a complex interplay among different neuronal populations \cite{koch1994large, sakurai1996population, quiroga2007decoding, vyas2020computation} \add{whose understanding is a central topic in systems neuroscience}. Several studies have investigated neural encoding at the level of individual neurons, showing that certain neurons selectively respond to specific features of incoming stimuli, such as spatial or temporal frequency, orientation, position, or depth \cite{kriegeskorte2021neural, barlow1967neural, campbell1968angular, henry1974orientation, olshausen1996emergence}. Due to advancements in the ability to simultaneously record the activity of large numbers of neurons across brain areas, recent decades have witnessed a shift of research focus towards the investigation of collective dynamics of neural populations \cite{gao2015simplicity, buzsaki2004large, brown2004multiple}. Remarkably, the trajectories of such populations are typically constrained in low-dimensional manifolds in the high-dimensional space of neural activity \cite{kriegeskorte2021neural, buonomano2009state, pandarinath2018inferring}, suggesting that the entire population dynamically encodes stimulus variables in this reduced \cite{cunningham2014dimensionality, mante2013context, gallego2017neural, remington2018flexible} or coarse-grained \cite{meshulam2019coarse, nicoletti2020scaling} neural state space. Tools from information theory have been used to measure the amount of information that the response of a neural system conveys on a stimulus \cite{brunel1998mutual, quian2009extracting, bernardi2015frequency, panzeri2022structures, mariani2022disentangling}, for instance through mutual or Fisher information \cite{shannon1948mathematical, blahut1987principles}. Yet, understanding how the emergent information properties depend on the underlying dynamics of the neural populations remains an open question.

Strong recurrent coupling and inhibition stabilization are common features of the cortex~\cite{sanzeni2020inhibition}. Crucially, such a finely balanced state not only prevents instability but may enhance the system's computational properties as well. Networks operating in this \add{fine-tuned} state often perform better in information processing tasks and complex computations \cite{langton1990computation, bertschinger2004real, beggs2008criticality}, while exhibiting optimal sensitivity to sensory stimuli \cite{kinouchi2006optimal, shew2009neuronal}. Such an interplay between strong excitatory coupling and compensatory inhibition is shaped by the connectivity structure between neural populations, which makes theoretical studies particularly challenging.

In this work, we explicitly tackle the problem of quantifying the information encoded by neuronal subpopulations on an external stochastic stimulus. By focusing on the connectivity between excitatory and inhibitory populations, \add{we compute the information between the neural activity and the external stimuli, providing analytical bounds on the mutual information in suitable limits. We} demonstrate that an excitatory-inhibitory balanced state is necessary not only to ensure stability \add{but also to maximize information encoding at the steady state, both in linear and nonlinear regimes}. Further, by studying the response to a single stochastic perturbation of varying intensity, we reveal an intrinsic trade-off between the optimal response at short and long times. In particular, global inhibition acts to regulate the total information encoded and the sensitivity of the system's response.

\begin{figure}[tb]
    \centering
    \includegraphics[width = \columnwidth]{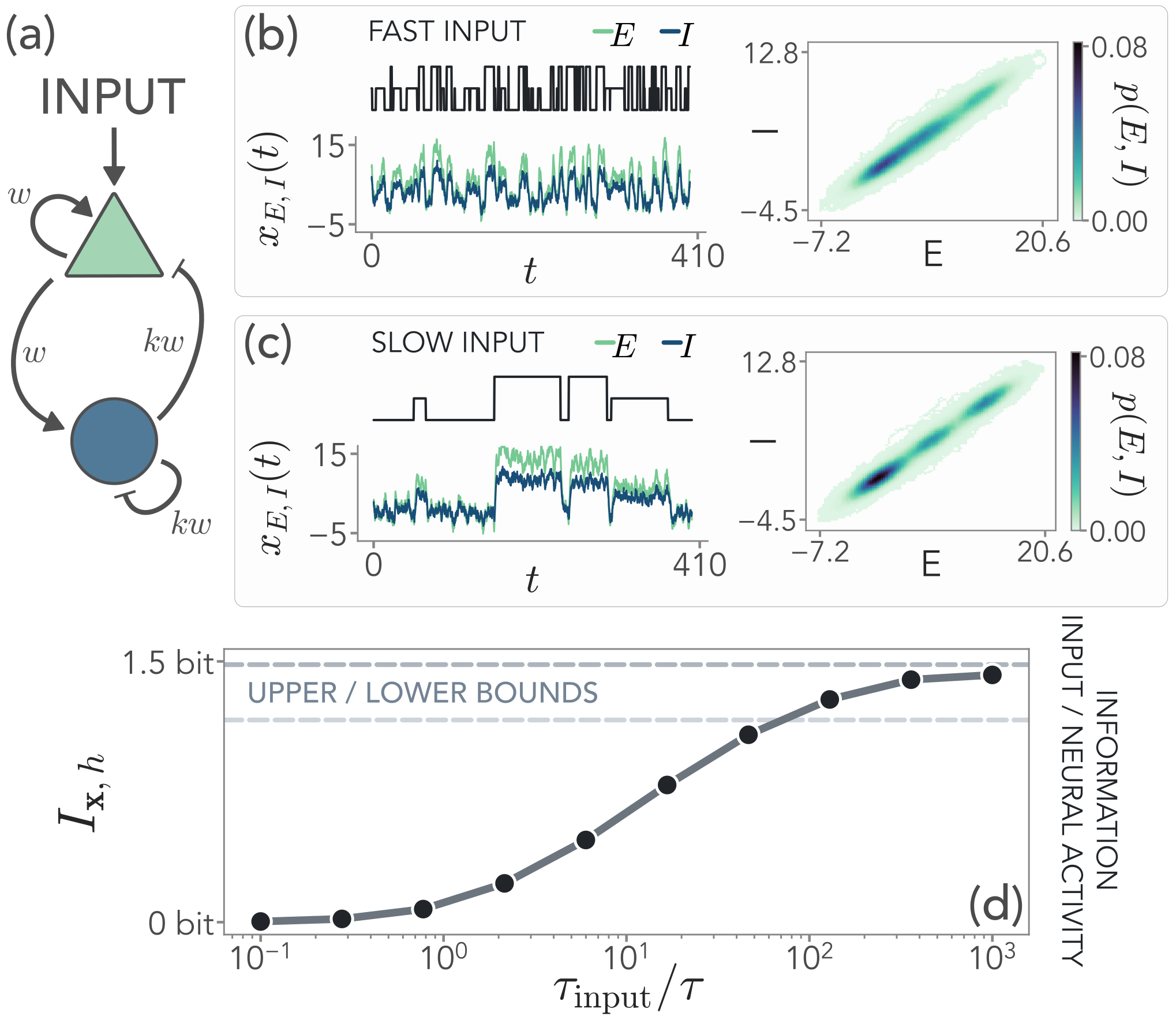}
    \caption{(a) Sketch of the model, describing a population of excitatory ($E$, green) and inhibitory neurons ($I$, blue) evolving on a timescale $\tau$. An input $h$ stimulates activity in the excitatory population on a timescale $\tau_\mathrm{input}$. (b-c) \add{In this figure, $f(x) = x$.} If $\tau_\mathrm{input} \ll \tau$, neural populations are not able to resolve different inputs. In the opposite limit, the joint probability $p(E, I)$ displays instead peaks around the different input strengths. (d) \add{$I_{\mathbf{x}, h}$ is evaluated by sampling numerically Eq.~\eqref{eq:langevin}. The mutual information} is zero in the fast-inputs regime, but sharply increases when $\tau_\mathrm{input} \gg \tau$, signaling that neural populations are capturing information on the input. \add{Parameters: $M=2$, $\tilde{q}_{\uparrow}=1/3$, $\tilde{q}_{\downarrow}=2/3$, $D=1/2$, $r=1$, $\tau=1$, $w=2$, $k=1.1$, $\Delta h = 2.5$ in all figures, unless stated otherwise. Analytical bounds are in Eq.~\eqref{eq:bounds}.}}
    \label{fig:figure1}
\end{figure}

To retain physical interpretability, we \add{consider the activity of two neuronal subpopulations, one excitatory $x_E$, and one inhibitory $x_I$.} The excitatory population receives a time-varying external input $h(t)$, representing the stimuli the neurons seek to encode in their dynamics, {\color{black}described by the Langevin equations
\begin{align}
\label{eq:langevin}
    \tau \frac{dx_{\mu}}{dt}= & - r_\mu x_\mu + \sum_{\nu \in {E,I}}A_{\mu\nu} f(x_\nu) + \\
    & + h(t) \delta_{\mu, E} + \sqrt{2 D_\mu \tau} \xi_\mu \nonumber
\end{align}
where $\tau$ is the characteristic neural timescale, $r_\mu$ is the decay of the activity, $\xi_\mu$ are independent white noises, and $f$ is an activation function. $A_{\mu\nu}$ is an element of the synaptic connectivity matrix
\begin{align}
    \hat{A} = \begin{pmatrix}
    w & -kw \\
    w & -kw
    \end{pmatrix} \, ,
\end{align}
with $w \ge 0$ and $k \ge 0$. Thus, $w$ measures the overall excitation strength, while $k$ quantifies the relative intensity of the inhibition. This model (Fig.~\ref{fig:figure1}a)} has been widely used in the literature as it captures essential properties of neuronal connectivity \cite{murphy2009balanced, schaub2015emergence, christodoulou2022regimes}. \add{In the Supplemental Material (SM) \cite{suppinfo_ref}, we also study the effect of different input projections by considering the case in which both populations receive the input.}

To model changes in the external environment, \add{the input follows a jump process between} a ground state $h_0$ and a set of $M$ environmental states $h_i$ \add{describing, e.g., sensory stimuli of different intensities, behavioral states, or motor commands}. For simplicity, we take $h_i = h_0 + i \Delta h$, with $h_0 = 0$ representing the absence of external signals \add{and $\Delta h$ is a constant shift between the inputs}. The \add{input switches from $h_0$ to any $h_i$} with uniform transition rates $q_{0 \to i}=q_{\uparrow}$, \add{and similarly} $q_{i \to 0}=q_{\downarrow}$. All other transition rates are set to zero, i.e., \add{no direct switches from one stimulus to another are present. Hence}, all environmental states are equally likely and the neural populations must respond to the stochastic jumps among them. The characteristic timescale of the \add{input} process is $\tau_\mathrm{input} = (q_{\uparrow} + q_{\downarrow})^{-1}$.

We seek to understand how much information the neuronal network can capture on the external inputs at stationarity. To this end, we compute the mutual information \cite{shannon1948mathematical, blahut1987principles}
\begin{align}
    I_{\boldsymbol{x},h}^\mathrm{st}
    &= \sum_{i=0}^M \int d\boldsymbol{x} \ p_{i,\boldsymbol{x}}^\mathrm{st} \log \dfrac{p_{i,\boldsymbol{x}}^\mathrm{st}}{p_{\boldsymbol{x}}^\mathrm{st}\pi_i^\mathrm{st}} \nonumber \\
    & = H_{\boldsymbol{x}} + H_\mathrm{input} - H_{\boldsymbol{x}, \mathrm{input}}
    \label{eq:mutual_information}
\end{align}
\add{where $H_{\boldsymbol{x}}$ is the differential entropy of the excitatory and inhibitory populations, $H_\mathrm{input}$ the Shannon entropy of the external inputs, and $H_{\boldsymbol{x}, \mathrm{input}}$ their joint entropy. Eq.~\eqref{eq:mutual_information} can be understood as} the Kullback-Leibler divergence between the joint steady-state probability of the inputs and the neural activity, $p^\mathrm{st}_{i,\boldsymbol{x}}$, with $\boldsymbol{x}=(x_E, x_I)$, and the corresponding marginal distributions, $p^\mathrm{st}_{\boldsymbol{x}}$ and $\pi_i^\mathrm{st}$ \cite{cover1999elements}. As such, $I_{\boldsymbol{x},h}^\mathrm{st}$ quantifies \add{all statistical dependencies between $\boldsymbol{x}$ and $h$ in terms of how much information is encoded in the joint probability of the input and the neural activity. For simplicity, we take} $D_\mu = D$ and $r_\mu = r$.

From Eq.~\eqref{eq:langevin}, the joint probability $p_{i,\boldsymbol{x}}(t)$ is governed by the Fokker-Planck equation:
\begin{align}
    \partial_t p_{i, \boldsymbol{x}}(t) =& \dfrac{1}{\tau} \sum_{\mu=E,I} \biggl[\partial_\mu [(\Tilde{F}_{i\mu}(\boldsymbol{x}) p_{i, \boldsymbol{x}}(t)]+ \partial_\mu^2 \state{i} \biggl] + \nonumber \\ 
    &+ \dfrac{1}{\tau_\mathrm{input}} \sum_{j=0}^M \biggl[ \tilde{q}_{j \to i} \state{j} - \tilde{q}_{i \to j} \state{i} \biggl]
    \label{eq:fokker_planck}
\end{align}
where $\tilde{q}_{j \to i}=\tau_\mathrm{input} q_{j \to i}$ are the rescaled transition rates, \add{$\Tilde{F}_{i\mu}(\boldsymbol{x})= -r x_\mu + \sum_{\nu} A_{\mu \nu}f(x_\nu) + h_{i(t)} \delta_{\mu,E}$} with $i(t)$ the environmental state at time $t$, and we used the shorthand notation $\partial_{x_\mu} := \partial_\mu$. Finding an explicit solution of Eq.~\eqref{eq:fokker_planck} is, in general, a formidably challenging task. However, exact solutions can be accomplished in a timescale separation limit~\cite{nicoletti2021mutual, nicoletti2022mutual, nicoletti2022information}.

In the limit of a fast-evolving input $\tau_\mathrm{input} \ll \tau$, the joint probability of the system factorizes as $p_{i,\boldsymbol{x}}^\mathrm{st}=p_{\boldsymbol{x}} ^\mathrm{st}\pi_i^\mathrm{st}$ (see SM \cite{suppinfo_ref}). \add{$\pi_i^\mathrm{st}$ is the stationary distribution of the inputs, and} $p_{\boldsymbol{x}}^\mathrm{st}$ is the solution of Eq.~\eqref{eq:langevin} with an effective input $\Tilde{h}_\mu = \delta_{\mu, E} \sum_i h_{\mu,i} \pi_i^\mathrm{st}$ (Fig. \ref{fig:figure1}b). In this regime, the mutual information between the neural populations and the input vanishes, i.e., $I_{\boldsymbol{x},h} \to 0$ when $\tau_\mathrm{input}/\tau \to 0$. Indeed, the stationary solution of the system shows that the neural activity is only influenced by the average input, as it cannot resolve its fast temporal evolution. On the other hand, in the limit of a slowly evolving external input $\tau_\mathrm{input} \gg \tau$, the system is described by the stationary probability $p_{i, \boldsymbol{x}}(t)=p_{\boldsymbol{x}|i}^\mathrm{st} \pi_i(t)$, where $p^\mathrm{st}_{\boldsymbol{x}|i}$ is the probability of the excitatory and inhibitory populations \add{at constant input} $h_i$. 

\add{We first focus on a linear activation function $f(x) = x$. In this case, the system is stable when $k > k_c = 1 - r/w$, while excitation is too strong for the inhibition to stabilize it for $k < k_c$. Then,} $p^\mathrm{st}_{\boldsymbol{x}|i}$ is a multivariate Gaussian distribution $\mathcal{N}(\boldsymbol{m}^\mathrm{st}_{i}, \hat{\Sigma}^\mathrm{st})$ with mean $\boldsymbol{m}^\mathrm{st}_{i}$
\begin{equation*}
\begin{pmatrix}
    {m}^\mathrm{st}_{E,i} \\ {m}^\mathrm{st}_{I, i} 
\end{pmatrix}
    =  \left(\hat{R} - \hat{A}\right)^{-1}
    \begin{pmatrix}
    h_i \\ 0
\end{pmatrix}
\end{equation*}
where $R_{\mu\nu} = r_\mu \delta_{\mu\nu}$, \add{and a covariance $\hat{\Sigma}^\mathrm{st}$ that} satisfies the Lyapunov equation:
\begin{equation*}
    \sum_{\nu} \left[\left(A_{\alpha\nu} - R_{\alpha\nu}\right) \Sigma^\mathrm{st}_{\nu\beta} + \Sigma^\mathrm{st}_{\alpha\nu} \left(A_{\nu\beta}^T -R_{\nu\beta}\right) \right]= -\add{2D_\alpha}\delta_{\alpha \beta}
\end{equation*}
as we show in the SM \cite{suppinfo_ref}. It is worth noting here that, since the input acts as an additional drift, it only changes the average of the distribution with respect to the case of no input. Therefore, the stationary probability distribution of the neural populations is the Gaussian mixture $p_{\boldsymbol{x}}^\mathrm{st} = \sum_i \pi_i^\mathrm{st} p_{\boldsymbol{x}|i}^\mathrm{st}$. We show a typical trajectory of the system and the corresponding probability distribution of neural activity in Fig.~\ref{fig:figure1}c. 

Even though the entropy of a Gaussian mixture cannot be written in a closed form, by employing the bounds proposed in \cite{kolchinsky2017estimating}, we obtain an upper and a lower bound on the mutual information starting from the Chernoff-$\alpha$ divergence and the Kullback-Leibler divergence between the mixture components (see SM \cite{suppinfo_ref}). We have that $I^{(b)}(\eta/4) \leq I_{\boldsymbol{x},h} \leq I^{(b)}(\eta)$, where
\begin{align}
    I^{(b)}(\eta) &= - \sum_{i=0}^M \pi_i^\mathrm{st} \log \biggl[ \sum_{j=0}^M \pi_j^\mathrm{st} e^{-(j-i)^2 \eta} \biggl]
    \label{eq:bounds}
\end{align}
with
\begin{align*}
    \eta = \frac{\Delta h^2}{4Dr} \dfrac{[r+w(k-k_c)] [2r^2+(3k-1)w + (k^2+1)w^2]}{w^2(k-k_c)(2r(k-k_c)+(k^2+1)w)} \; .
\end{align*}
Since $I^{(b)}(\eta/4) > 0$, we have that $I_{\boldsymbol{x},h}$ is always non-zero. Eq.~\eqref{eq:bounds} shows that, in the limit of a slow input, the excitatory and inhibitory populations are able to capture information on the external stimulus. In the intermediate regime between the fast- and slow-input limits, we cannot solve the Fokker-Planck equation explicitly. However, a direct simulation of the system shows that the mutual information smoothly interpolates between the two regimes, as we see in Fig.~\ref{fig:figure1}d. Taken together, our results underscore the significance of timescales for neuronal circuits and their capability of processing information on external time-varying stimuli \cite{butts2007temporal, das2019critical, mariani2021b, nicoletti2024information, nicoletti2024gaussian}.

\begin{figure}[tb]
    \centering
    \includegraphics[width = \columnwidth]{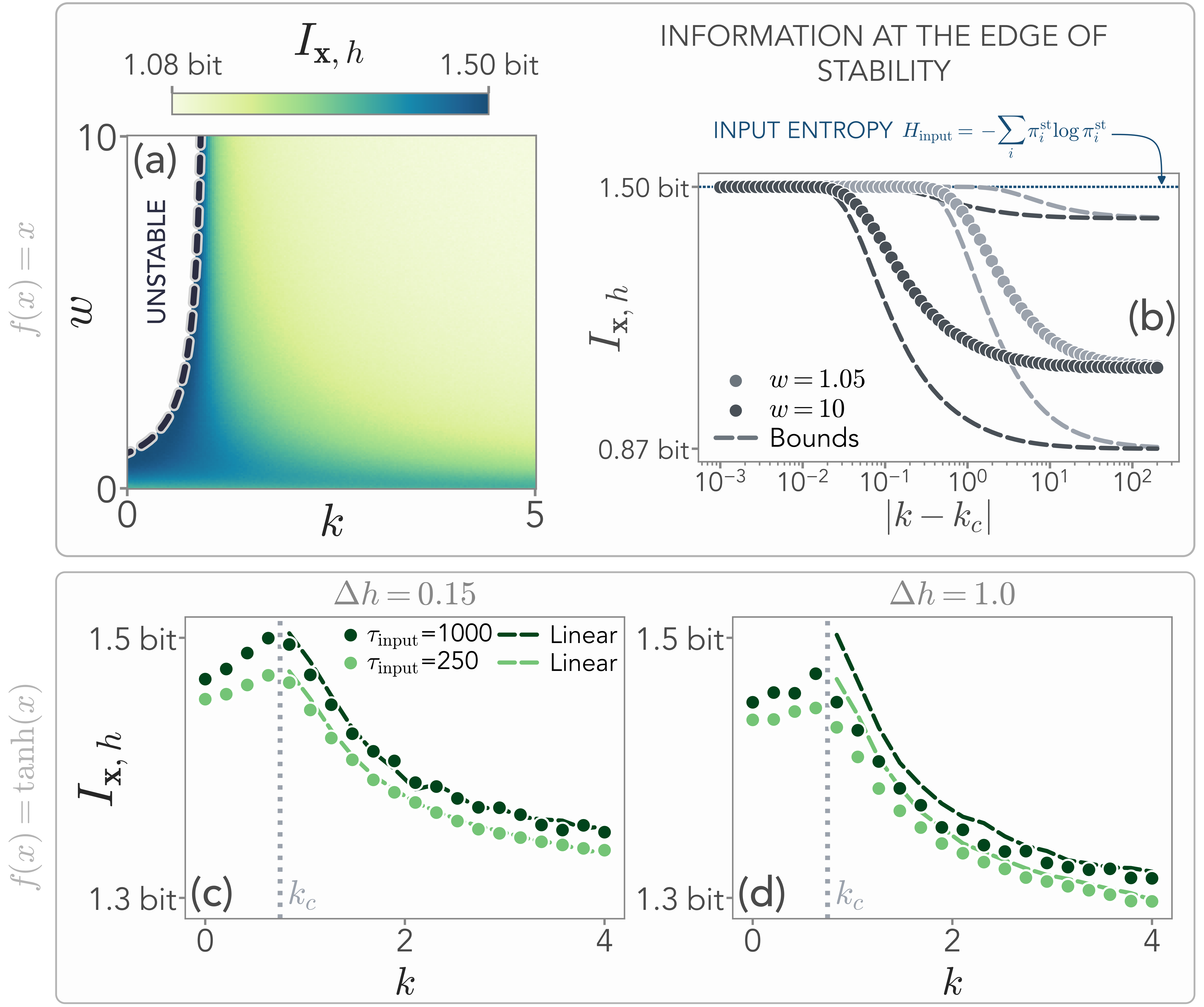}
    \caption{\add{(a-b) Mutual information between the neural populations and the input in the linear case. If $k < k_c$ (black dotted line), the system is unstable. Information is maximized at the edge of stability, $k \to k_c$. In this limit, $I_{\boldsymbol{x},h}$ converges to $H_\mathrm{input}$, which quantifies the information contained in the input. Mutual information is obtained from numerically integrating the Gaussian mixture, while analytical bounds are in Eq.~\eqref{eq:bounds}. (c-d) Comparison with the nonlinear case. For small $\Delta h$, in the linear stability regime, results are comparable, whereas they differ for larger $\Delta h$. $I_{\boldsymbol{x},h}$ peaks at $k \approx k_c$, and decreases when $k < k_c$. Mutual information is obtained numerically with a kNN estimator. 
    In (c), $D=0.001$, while $D=0.05$ in (d) to improve the numerical estimations.}}
    \label{fig:figure2}
\end{figure}

\begin{figure*}[tb]
    \centering
    \includegraphics[width = \textwidth]{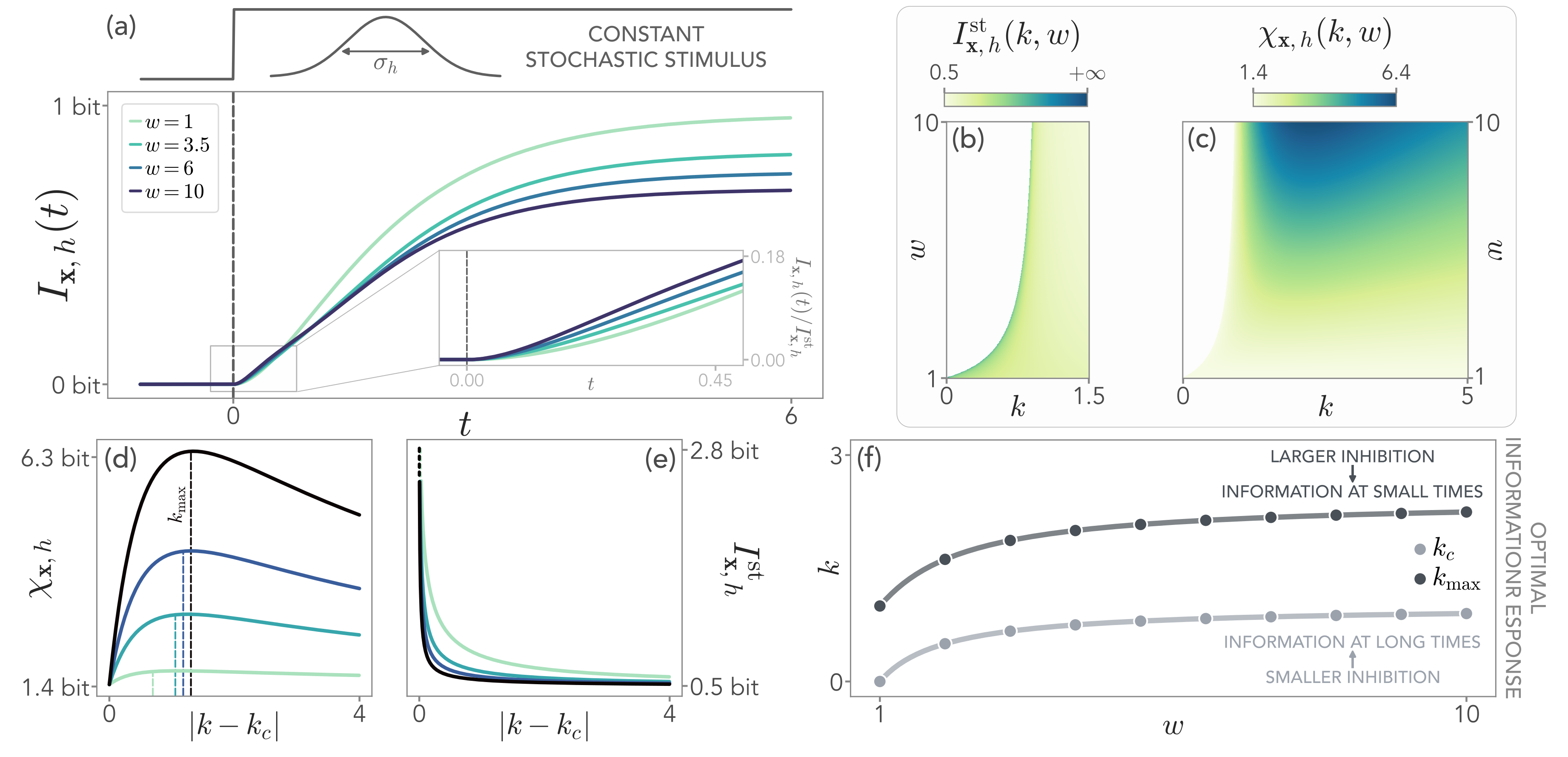}
    \caption{(a) Dynamics of mutual information \add{with a constant stochastic input ($\sigma_h = 1$)}. (b-c) \add{$I_{\boldsymbol{x},h}^\mathrm{st}$} \add{(Eq.~\eqref{eq:mutual_information_time_dep_gaussian})} diverges at the edge of stability, \add{whereas} $\chi_{\boldsymbol{x},h}$ \add{(Eq.~\eqref{eq:chi})} peaks at intermediate values of $k$ for large $w$. (d-e) Sensitivity peak occurs for $k_\mathrm{max}(w) > k_c$, while $I_{\boldsymbol{x},h}^\mathrm{st} \to +\infty$ for $k \to k_c$. (f) Thus, a larger inhibition strength $k$ benefits the \add{short-time response when the input arrives}. On the contrary, at long times, information is maximized by reducing $k$ and approaching the edge of stability at $k = k_c$.  
    }
    \label{fig:figure3}
\end{figure*}

Crucially, the synaptic strengths of the excitatory and inhibitory populations drastically affect their mutual information with the input. Indeed, as we show in Fig.~\ref{fig:figure2}a, $I_{\boldsymbol{x},h}$ strongly depends on the interplay between excitation and inhibition. Furthermore, the bounds in Eq.~\eqref{eq:bounds} tighten as $k$ approaches $k_c$, eventually collapsing to one single value in the limit $k \to k_c$, which corresponds to the edge of stability of the system (see Fig.~\ref{fig:figure2}b):
\begin{equation}
    I_{\boldsymbol{x},h} \underset{k \to k_c}{\longrightarrow} H_\mathrm{input} = - \sum_{i = 0}^M \pi_i^\mathrm{st}\log \pi_i^\mathrm{st} \;.
    \label{eq:mutual_limit}
\end{equation}
Eq.~\eqref{eq:mutual_limit} tells us that, at the edge of stability, the neural populations are able to fully capture the information contained in the external input, which is exactly its entropy $H_\mathrm{input}$. Intriguingly, we also find that this is the maximum value the mutual information can attain, as shown in Fig.~\ref{fig:figure2}b. Thus, modulation of the inhibition strength plays a prime role in determining how efficiently the system can encode the external inputs, and the corresponding mutual information sharply increases as the edge of stability is approached. \add{Notably, $\eta$ diverges also when $\Delta h^2/D \to \infty$, leading to maximal mutual information $H_\mathrm{input}$. This shows how the information encoded in neural activity crucially depends on the relative strength of input and noise (see SM \cite{suppinfo_ref}).}

\add{Then, we consider a more realistic scenario of a nonlinear activation function $f(x) = \tanh(x)$. Since no analytic expression exists for $p_{\boldsymbol{x}|i}^\mathrm{st}$, we estimate $I_{\boldsymbol{x},h}$ numerically from the Langevin trajectories using a kNN estimator \cite{kraskov2004estimating}. In Fig.~\ref{fig:figure2}c, we show that for small $\Delta h$ the results are similar to the linear case in the stable region, whereas for larger $\Delta h$ we find quantitative differences between the two due to saturation effects (see Fig.~\ref{fig:figure2}d and SM \cite{suppinfo_ref}). Crucially, in the presence of a nonlinear activation function the mutual information peaks at $k \approx k_c$ and decreases beyond the region of linear stability. These observations hint at the robustness of our results outside the linear scenario.}

So far, we have considered the steady-state response of the system to a time-varying input. However, understanding how \add{and how quickly} neural populations dynamically acquire information \add{when a single persistent input is presented} is crucial as well. \add{We now consider a system in its stationary state that is} perturbed by an external stochastic input. In this case, the mutual information in time reads
\begin{align}
    I_{\boldsymbol{x},h}(t)
    = H_{\boldsymbol{x}}(t) - \int_{-\infty}^{\infty} dh \, p_h^\mathrm{st}(h) \, H_{\boldsymbol{x}|h}(t)
    \label{eq:mutual_information_time_dep}
\end{align}
where $p_h^\mathrm{st}$ is the distribution of the strength of the stimulus, and $ H_{\boldsymbol{x}|h}$ is the conditional differential entropy of $\boldsymbol{x}$ at a given input value $h$. We assume that $p_h^\mathrm{st}$ is fully characterized by the mean input strength, $\mu_h$, and its variance, $\sigma_h$, so that $p_h = \mathcal{N}(\mu_{h}, \sigma_{h})$. Then, we have:
{\color{black}
\begin{align}
    I_{\boldsymbol{x},h}(t) 
    = \dfrac{1}{2} \log \dfrac{\det\left[\hat{\Sigma}^\mathrm{st} + \hat{K}(t) \begin{pmatrix}
    \sigma_h^2 & 0 \\
    0 & 0
    \end{pmatrix}
     \hat{K}(t)^T\right]}{\det\left(\hat{\Sigma}^\mathrm{st}\right)}
     \label{eq:mutual_information_time_dep_gaussian}
\end{align}
where} $\hat{K}(t)$ is the time-dependent gain matrix that we derive in the SM \cite{suppinfo_ref}. In Fig.~\ref{fig:figure3}a, we plot the time evolution of the mutual information. At long times, we find once more that the mutual information is maximized at the edge of stability (Fig.~\ref{fig:figure3}b), with $I_{\boldsymbol{x},h}$ diverging as $k \to k_c$. We note that, since the differential entropy for the continuous input distribution is not necessarily positive, the bounds in Eq.~\eqref{eq:bounds} cannot be straightforwardly applied. In particular, while the maximal information content of the input was associated with its switching dynamics in the previous setting, there is now no a priori limit to the information that the system can encode.

The scenario becomes more intricate at short times after the stimulus. In the inhibition-stabilized regime, where $w>1$, the response of the neural populations exhibits a faster increase for stronger excitatory couplings away from the edge of instability. To assess the system's responsiveness, we introduce a metric of sensitivity{\color{black}:
\begin{align}
    \label{eq:chi}
    \chi_{\boldsymbol{x},h} & = \frac{\partial^2 I_{\boldsymbol{x},h}(t)}{\partial t^2} \biggr|_{t=t_\mathrm{stim}} \\
    & = \frac{\sigma_h^2}{2 D}\frac{[r+w(k-k_c)][2w k_c^2 +r(k+1)]}{2r(k - k_c) + w(1 + k^2)} \nonumber
\end{align}
which measures how quickly the information on the stimulus increases immediately after the stimulation time $t=t_\mathrm{stim}$} \cite{derivative_note}. In Fig.~\ref{fig:figure3}c, we show that $\chi_{\boldsymbol{x},h}$ peaks at an optimal inhibition strength \add{$k_{\rm max}(w)>1$}, whose complete expression is given in the SM \cite{suppinfo_ref}. Crucially, this optimal value depends on the excitation strength (see Fig.~\ref{fig:figure3}d). This \add{reveals} that the inhibition regime for the optimal response at short times is drastically different from that at long times, \add{as we show in Figs.~\ref{fig:figure3}d-e}. In particular, since $k_c (w) < k_\mathrm{max}(w)$ for all $w$, our results unravel a fundamental trade-off between achieving maximum accuracy and the speed at which the neural populations encode information about the external stimulus, akin to a speed-accuracy trade-off \add{emerging in different biological contexts} \cite{lan2012energy}. \add{Remarkably, similar trade-offs have been recently found by studying the response of recurrent neural networks with random connectivities at different observation times \cite{azizpour2023available}.} As we show in Fig.~\ref{fig:figure3}f, long-time accuracy is generally achieved at lower inhibition, whereas sensitivity maximization requires a larger value of $k$. {\color{black}In the SM \cite{suppinfo_ref}, we show that, when both populations receive the prolonged stimulus, sensitivity is maximized at $k\to \infty$, suggesting that inhibition-dominated networks exhibit an enhanced tracking ability of the input \cite{renart2010asynchronous}.}

Overall, our analysis marks a significant step towards the understanding of how the structure of connectivity shapes information encoding in neuronal population dynamics. We \add{have shown, both} analytically in an exactly solvable regime \add{and numerically for a nonlinear scenario,} that the mutual information between a \add{switching} input and the receiving neuronal populations is \add{controlled by the balance between the excitatory and inhibitory couplings, and peaks at the edge of linear stability. Moreover, we found that an increased inhibition strength is instrumental for establishing a robust response at short times,} highlighting the importance of \add{precisely tuning excitation and inhibition to achieve optimal encoding at different timescales}. As non-normal synaptic interactions are crucial for \add{realizing} this optimal state, \add{our findings at a coarse-grained level underscore the essential role of the underlying connectivity. In particular, structural connectivity has been shown to be} essential in supporting complex dynamical evolution both in whole-brain connectomes \cite{barzon2022criticality} and in artificial recurrent neural networks \cite{baggio2020efficient}. \add{Our study opens the avenue for an information-theoretic quantification of these emerging features from first principles, at the level of neural populations. In future works, the extension of similar ideas to more microscopic models might shed light on the intimate link between complex, even chaotic \cite{langton1990computation, clark2024theory}, dynamics, and information processing performances.}

Notably, alterations in excitatory-inhibitory balance have been experimentally related to the loss of information-processing efficiency observed in pathological conditions \cite{dehghani2016dynamic, sohal2019excitation}. Our predictions are also consistent with recent experimental studies in which theoretical tools from response theory have been applied to extensive whole-brain neuronal recordings. The emergent dynamics of several brain regions has been shown to lie at the edge of stability~\cite{dahmen2019second, morales2023quasiuniversal}, with a distance from instability that only slightly varied along the cortex. Such heterogeneity might be explained as an increase in the inhibition level \cite{wang2020macroscopic}, \add{and} our findings suggest that this observed feature may be related to the tuning of sensitivity to different timescales \cite{murray2014hierarchy, manea2022intrinsic}. Importantly, the external \add{input} considered here may be immediately generalized to a high-dimensional signal representing, for example, multiple stimuli with different characteristics (e.g., frequency, intensity) targeting spatially separated populations \add{potentially evolving on different timescales}.

Although we focused on a paradigmatic - yet widely used - model, our approach can be \add{extended} to investigate more detailed \add{and microscopic} synaptic structures and \add{include the role played by plasticity to drive accurate encoding.} \add{Overall,} our work paves the way to the unraveling of the fundamental mechanisms supporting information encoding and sensitivity in neuronal networks.

\begin{acknowledgments}
\emph{Acknowledgments}.---G.N. acknowledges funding provided by the Swiss National Science Foundation through its Grant CRSII5\_186422.
\end{acknowledgments}


\begin{thebibliography}{10}
\bibitem{derivative_note}
We use the second derivative since the first derivative is always zero at $t=t_{\rm stim}$.
\end{thebibliography}

\end{document}